\documentclass[prl, showpacs, twocolumn, floatfix]{revtex4}

\usepackage{graphicx}
\usepackage{amsmath, amsfonts, amssymb, bm}
\usepackage[]{psfrag}

\psfragscanoff
\usepackage{pifont}
\usepackage{dcolumn}% Align table columns on decimal point
\usepackage{slashed}

\begin{document}

\title{Relativistic three-body recombination with the QED vacuum}

\author{Huayu Hu$^{1,2}$}
\author{Carsten M\"uller$^1$}
\affiliation{$^1$Max-Planck-Institut f\"ur Kernphysik, Saupfercheckweg 1, 69117 Heidelberg, Germany\\
$^2$Department of Physics, National University of Defense Technology, Changsha 410073, P. R. China}

\date{\today}

\begin{abstract}
Electron-positron pair annihilation into a single photon is studied when a second free electron is present. Focussing on the relativistic regime, we show that the photon emitted in the three-lepton interaction may exhibit distinct angular distributions and polarization properties. Moreover, the process can dominate over two-photon annihilation in relativistic electron-positron plasmas of few-MeV temperature. An analogy with three-body recombination of electrons with ions is drawn.
\end{abstract}

\maketitle
{\emph{Introduction.}---}
Electron-positron annihilation into photons is a fundamental process of quantum electrodynamics (QED) which allows for various applications \cite{PAS}. 
In free space, energy-momentum conservation dictates electrons and positrons to annihilate into at least two photons. While the two-photon process $e^+e^-\to 2\gamma$ is most abundant, larger photon numbers may also be emitted, as the three-photon decay of ortho-positronium shows. 

Annihilation into a single photon is usually forbidden kinematically. It becomes possible, however, in the presence of an additional particle or external field which can absorb recoil momentum. Such a process has been studied most intensively, both in experiment and theory, for positron annihilation on deeply bound inner-shell electrons in atoms \cite{boundTheo}. Single-photon annihilation can also occur in the presence of a (quasi) free spectator electron (or positron) via 
\begin{eqnarray}
e^+e^- + e \to \gamma + e'\,.
\label{SPA}
\end{eqnarray}
The low-energy limit of this reaction has been studied with respect to the decay of positronium ions Ps$^-$ where the second electron is loosely bound to the Ps core. Since the calculation is quite involved, the first studies \cite{Lynn} treated only subsets of the relevant leading-order Feynman diagrams. Later on, a complete consideration was provided \cite{Ps8}. For the ratio of single-photon to two-photon annihilation in Ps$^-$, the small value $R_{1\gamma}/R_{2\gamma}\sim 10^{-10}$ was found, which cannot be probed in experiment yet \cite{PsExp}. Contrary to that we shall show in the present study that, in the domain of high energies, reaction \eqref{SPA} becomes significant
and acquires very special features.

Relativistic $e^+e^-$ samples of high density ($\rho\sim 10^{16}$\,cm$^{-3}$) can nowadays be produced via intense laser-solid interactions \cite{Chen}. A further increase in density up to $\rho\sim 10^{23}$\,cm$^{-3}$ by this method has been predicted \cite{Meyer-ter-Vehn}. Similar pair densities are expected from QED cascades occurring in collisions of superintense laser beams \cite{Ruhl}. This offers prospects for laboratory studies on relativistic $e^+e^-$ plasmas which are of relevance to various astrophysical phenomena such as gamma-ray bursts, supernova explosions, and the evolution of the early universe \cite{astronomy}. Meanwhile, theoreticians have started to explore
(ultra-) relativistic $e^+e^-$ plasmas with densities $\gtrsim 10^{30}$\,cm$^{-3}$, analyzing the thermalization process \cite{Ruffini}, dynamical properties \cite{Thoma}, and photo-emission spectra  \cite{Kaempfer}. In particular it was shown that in these extremely dense environments higher-order QED processes and multi-particle correlation effects may be prominent.

In this Letter, the single-photon annihilation \eqref{SPA} is studied in the relativistic regime. We develop an alternative approach to the problem based on Furry-Feynman diagrams and show that the relative importance of this process can grow for higher particle energies. It is found to dominate over the usual two-photon annihilation channel in a relativistic $e^+e^-$ plasma above 3 MeV temperature. Characteristic features in the angular distribution and polarization properties of the emitted photon are revealed. Besides, a connection to atomic physics is established by relating the triple reaction \eqref{SPA} with three-body recombination of electrons with ions.

\begin{figure}\centering
\includegraphics[height=4cm,width=7.5cm]{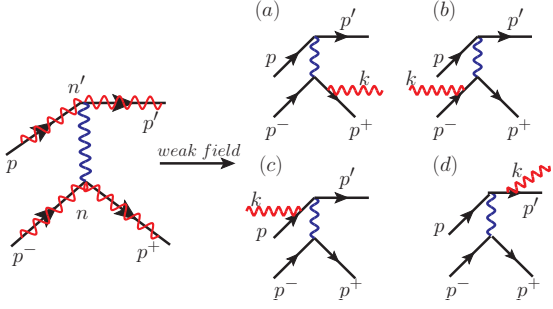}
\caption{\label{Fig:sppanni:Ffeynman} (Left) Furry-Feynman diagram of $e^+e^-$ pair annihilation upon scattering off another electron in the presence of an external plane-wave photon field. The zigzag-lines represent field-dressed (Volkov) states; they are labeled by the particle momenta outside the field. $n$ and $n'$ are the numbers of real photons absorbed at each vertex, respectively. (Right) In the limit of a weak external field, the Furry-Feynman diagram can be expanded to leading order into four ordinary Feynman graphs (a)-(d). The corresponding exchange Furry-Feynman diagram (with $p^- \leftrightarrow p$) accounts in the same manner for the remaining four field-free Feynman graphs.}
\end{figure}

{\emph{Theoretical approach.}---} Calculation of \eqref{SPA} by the methods of ordinary QED requires consideration of eight Feynman diagrams to leading order, four of which are shown in  Fig.\,\ref{Fig:sppanni:Ffeynman}. The other four result from exchange of the incoming electron lines. 
Treating the problem instead within the framework of laser-dressed QED reduces its diagrammatic complexity. Within this approach leptons are described by Volkov states \cite{Landau} which include the interaction with an external plane-wave electromagnetic field to all orders (Furry picture). It usually serves to calculate nonperturbative multiphoton processes in atomic physics and QED (see \cite{ldQED} for a review), but may also be utilized to obtain single-photon cross sections: In an expansion of the laser-dressed matrix element with respect to the laser-particle coupling parameter, the leading terms recover the coherent sum of the leading-order ordinary Feynman diagrams when the amplitude of the laser field is renormalized appropriately (see chapter 101 in \cite{Landau} for an example regarding Compton scattering).

In the present case, the eight ordinary Feynman graphs describing reaction \eqref{SPA} can be obtained from the weak-field limit of just two Furry-Feynman diagrams, as indicated in Fig. \ref{Fig:sppanni:Ffeynman}. Using relativistic units with $\hbar=c=1$, the single-photon annihilation amplitude may thus be written as
\begin{align}
S=&i2(2\pi)^5\alpha\delta^4(p^++p^-+p-k-p')\sqrt{\frac{2\pi m^4}{E^+ E^- E_p E_{p'} \omega V^5}}\,\nonumber\\
&\times(\mathcal{M}_{p^+p^-,pp'}-\mathcal{M}_{p^+p,p^-p'})\,,
\end{align}
where $\alpha=e^2\approx 1/137$ is the QED fine-structure constant, $m$ the electron mass, $V$ the interaction volume, $E^+$, $E^-$, $E_p$, $E_{p'}$ and $\omega$ are the particle and photon energies, and $\mathcal{M}_{p^+p^-,pp'}$ is the normalized matrix element corresponding to the weak-field limit of the Furry-Feynman diagram shown in Fig. \ref{Fig:sppanni:Ffeynman}. It is given by
\begin{equation}\label{Eq:M_fianni}
\mathcal{M}_{p^+p^-,pp'}=\displaystyle\sum_{n=0,1} \frac{ M^\mu(p^+,p^-|n) N_\mu(p,p'|n')}{(p^++p^--nk)^2}
\, ,
\end{equation}
with $n$ being the number of photons emitted by the annihilating pair of particles, and $n'=1-n$ being the number of photons emitted by the spectator particle.
Moreover,
\begin{align}
& M^\mu=\bar{v}_{p^+,s^+}\left[b_n\gamma^\mu-\left(-\frac{e\slashed{\epsilon}\slashed{k}\gamma^\mu}{2k\cdot p^+}
+\frac{e\gamma^\mu\slashed{k}\slashed{\epsilon}}{2k\cdot p^-}\right)
c_n\right]u_{p^-,s^-}\,,\nonumber\\
& N_\mu=\bar{u}_{p',s'}\left[B_{n'}\gamma_\mu-\left(\frac{e\slashed{\epsilon}\slashed{k}\gamma_\mu}{2k\cdot p'}
+\frac{e\gamma_\mu\slashed{k}\slashed{\epsilon}}{2k\cdot p}\right)
C_{n'}\right]u_{p,s}\,,
\end{align}
where $u$, $v$ and $\gamma^\mu$ are the usual Dirac spinors and matrices, respectively, and $\epsilon$ is the polarization four-vector of the emitted photon. The leading-order values of the coefficients are
\begin{align}
& b_0=1\,,c_0=0\,,b_1=\frac{e(\epsilon\cdot p^-)}{k\cdot p^-}-\frac{e(\epsilon\cdot p^+)}{k\cdot p^+}\,,c_1=1\,,\nonumber\\
&
B_0=1\,,C_0=0\,,B_1=\frac{e(\epsilon\cdot p)}{k\cdot p}-\frac{e(\epsilon\cdot p')}{k\cdot p'}\,,C_1=1\,.
\end{align}
The quantity $\mathcal{M}_{p^+p,p^-p'}$ differs from $\mathcal{M}_{p^+p^-,pp'}$ by the exchange of the momenta of the two incoming electrons $(p^-\leftrightarrow p)$. The total rate is obtained as
\begin{equation}
R_{1\gamma}=\frac{1}{\mathcal{T}} \int\frac{Vd^3p'}{(2\pi)^3} \int\frac{Vd^3k}{(2\pi)^3} \frac{1}{8}\sum_{\rm pol's}\sum_{\rm spins} |S|^2\,, 
\label{Rtot}
\end{equation}
where $\mathcal{T}$ is the interaction time, and a statistical factor $\frac{1}{8}$ comes from the initial spin-state averaging. 
Note that $R_{1\gamma}\propto V^{-2}$ exhibits a different density dependence than the rate for two-photon annihilation. In the low-energy limit, it amounts to 
$R_{1\gamma}=2^5 \pi^2 \alpha^3/(3^3 m^5 V^2)$ \cite{Lynn,Ps8}.

It is convenient to define an annihilation rate per positron, 
\begin{equation}
 R_{1\gamma}^{e^+}=N_pN_{p^-}R_{1\gamma}=\rho_p\rho_{p^-}V^2R_{1\gamma}\,,
\label{Rpos}
\end{equation}
where $N_p=\rho_pV$ (and accordingly $N_{p^-}=\rho_{p^-}V$) is the number of electrons with momentum $p$ in the volume $V$. 

Based on Eq.\,\eqref{Rpos} we will now discuss the properties of reaction \eqref{SPA} in the domain of relativistic energies.

{\emph{Annihilation in the collision of a relativistic electron with a low-energy $e^+e^-$ pair.}---} In order to gain detailed insights into reaction \eqref{SPA} we first study it in a special configuration. Let us assume that a high-energy electron beam (of energy $E_p$) penetrates a cold $e^+e^-$ sample (with energies $E^+\approx E^-\approx m$) \cite{Cassidy,Coulomb-effects}.

\begin{figure}[b]
\begin{center}
\includegraphics[height=4cm,width=7cm]{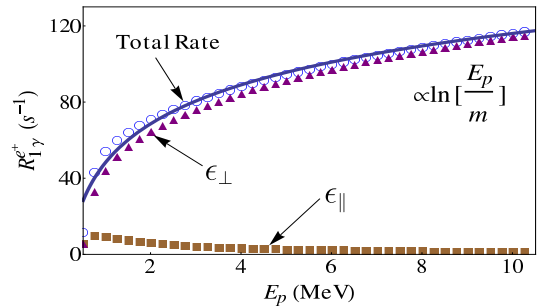}
\caption{\label{Fig:anniEp} Total rate for annihilation of an $e^+e^-$ pair at rest into a single photon, as a function of the energy of the second electron [see Eq.\,\eqref{SPA}]. The particle densities are chosen as $\rho_p=\rho_{p^-}=10^{24}\,{\rm cm}^{-3}$. The solid line is a logarithmic fit function to the numerical results shown by the open circles. Also shown are the separate contributions to the rate stemming from photons with polarization vector perpendicular ($\epsilon_\perp$) or parallel ($\epsilon_\parallel$) to the plane spanned by ${\bf k}$ and ${\bf p}$.}
\end{center} 
\end{figure}

As Fig.\,\ref{Fig:anniEp} shows, the single-photon annihilation rate increases with the incident electron energy $E_p$. To a good approximation, the rate follows a logarithmic increase
\begin{equation}
R_{1\gamma}^{e^+}\approx   \frac{3\alpha^3\pi^2}{m^5}\rho_p\rho_{p^-}\left[\ln\left(\frac{E_p}{m}\right)+1\right]\,.
\end{equation}
One can show that, in the case $E_p\gg m$, the rate is dominated by the contributions from the diagrams $(a)$ and $(b)$  in Fig. \ref{Fig:sppanni:Ffeynman}. They describe channels where the photon is emitted by the low-energy electron or positron and give rise to the logarithmic growth of the rate. The latter results from the fact that the intermediate photon approaches the mass shell as $E_p$ increases, according to $k^2_{\rm vir}=(p^++p^--k)^2\approx\frac{m^4}{E_p^2}$. 
The contribution from the other diagrams, describing photoemission by the colliding electron or annihilation of the colliding electron with the positron, is suppressed.
Thus, for large $E_p$, the process \eqref{SPA} may be viewed approximately as being composed of two successive processes: two-photon annihilation followed by the reverse process of electron radiation. A small ``virtuality'' of the intermediate photon must clearly remain, as the latter process is forbidden for a real photon.
Note besides that despite the small value of $k^2_{\rm vir}$ the three-momentum $|\mathbf{k}_{\rm vir}|\approx m$ is large. Hence $\rho\ll |\mathbf{k}_{\rm vir}|^3$, so that collective effects (such as the Landau-Pomeranchuk-Migdal effect which would restrict the coherence length of the process) are negligible.

Particularly interesting is the polarization of the emitted photon. As Fig.\,\ref{Fig:anniEp} shows, in the regime $E_p\gg m$ the photon is largely polarized along the direction $\epsilon_\perp$ perpendicular to the plane spanned by the photon wave vector ${\bf k}$ and the colliding electron momentum ${\bf p}$. An intuitive explanation of this finding may be given based on the dominance of the diagrams $(a)$ and $(b)$, together with the fact that the intermediate photon is almost "real". First we note that in classical electrodynamics the radiation of an accelerated charge of high velocity is polarized in the plane spanned by the radiation wave vector and the charge velocity vector \cite{ED}. Besides it is known that, due to selection rules, the polarization planes of the photons generated in two-photon annihilation of slow electrons and positrons are orthogonal \cite{Wheeler}. By combining these two arguments we may infer that in our case the nearly on-shell intermediate photon absorbed during scattering of the high-energy electron will be polarized preferentially along $\epsilon_\parallel$, whereas the real photon emitted in the annihilation process is polarized along $\epsilon_\perp$ indeed \cite{diminish}.

\begin{figure}\centering
\includegraphics[height=4cm,width=7cm]{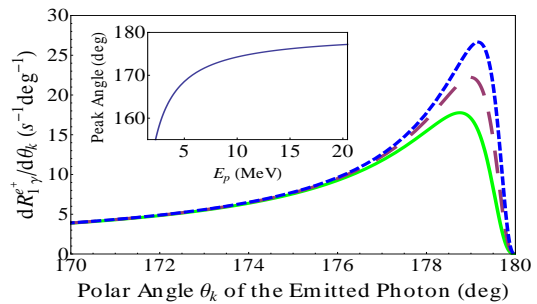}
\caption{\label{Fig:anniangularpeak} Angular distribution of the photon emitted in reaction \eqref{SPA} for $E^+=E^-\approx m$ and various $E_p$ values: $E_p=40$\,MeV (solid line), $E_p=50$\,MeV (long-dashed line), $E_p=60$\,MeV (short-dashed line). The angle $\theta_k$ is measured with respect to the momentum vector ${\bf p}$. The inset shows the $E_p$ dependence of the angle where the emission maximum lies.}
\end{figure}

For $E_p\gg m$, the energy spectrum of emitted photons $dR^{e^+}_{1\gamma}/d\omega\propto[E_p(\omega-m)]^{-2}$
sharply peaks at the smallest possible photon energy $\omega\approx m(1+\frac{m^2}{4 E_p^2})$. The angular photon distribution, see Fig. \ref{Fig:anniangularpeak}, is well described by
\begin{align}\label{anni:angular}
 \frac{dR^{e^+}_{1\gamma}}{d\theta_k}
\approx&\frac{\alpha^3\pi^2E_p\rho_p\rho_{p^-}\sin^3\theta_k(1-\cos\theta_k)}{2m^6(1+\frac{1-\cos\theta_k}{2m}E_p)
(1+\cos\theta_k)^2}\,,
\end{align}
where $\theta_k=\angle({\bf k},{\bf p})$. The photon is preferably emitted close to the backward direction of the colliding electron. This can be understood by observing that the radiation of a relativistic electron is directed mainly along its direction of motion \cite{ED}. Thus, in the time-reversed process, the electron preferably absorbs the virtual photon when its wave vector $\mathbf{k}_{\rm vir}$ is almost parallel to $\mathbf{p}$. Since $\mathbf{k}_{\rm vir}+\mathbf{k}= 0$, the wave vector of the emitted photon $\mathbf{k}$ should mainly lie in the opposite direction. Photo-emission exactly along $-\mathbf{p}$, however, is forbidden by the four-momentum conservation condition.

{\emph{Annihilation in a relativistic $e^+e^-$ plasma.}---} We now turn to a situation where reaction \eqref{SPA} becomes comparable with two-photon annihilation in terms of total rates. 

Let us consider a relativistic $e^+e^-$ plasma which is homogeneous, isotropic, and in its thermal and chemical equilibrium, with equal $e^+$ and $e^-$ densities. The single-photon annihilation rate per volume in the plasma, $R_t$, may be estimated by convoluting the rate of Eq.\,\eqref{Rtot} with the Fermi-Dirac distributions $n_F(E,T)=\frac{g_F}{e^{(E-m)/T}+1}$ of the particles, where $T$ is the plasma temperaturee and $g_F=2$ the number of spin degrees of freedom:

\begin{align}\label{annitotalrate}
R_t(T)=&\frac{2S}{V}\int\frac{Vd^3p^+}{(2\pi)^3} n_F(E^+,T)\int\frac{Vd^3p^-}{(2\pi)^3} n_F(E^-,T)\nonumber\\&\times\int\frac{Vd^3p}{(2\pi)^3} n_F(E_p,T)  R_{1\gamma}(p^+,p^-,p)\,.
\end{align}
Here, $S=\frac{1}{2}$ is a statistical factor to avoid double counting of the two electrons involved, and the factor 2 appears because -- apart from \eqref{SPA} -- there is a second symmetric single-photon process: $e^+e^-+ e^{+'}\to e^{+''} + \gamma$. Note that, although the plasma density $\rho$ may be very high, a perturbative QED treatment of the particle interactions is rendered applicable by the small value of the Coulomb coupling parameter $\approx\alpha\rho^{1/3}/T\ll 1$ which measures the ratio of potential to kinetic particle energies. However, relying solely on individual particle collisions, Eq.\,\eqref{annitotalrate} does not account for collective effects due to long-range interactions with the plasma medium \cite{Thoma,Kaempfer}.

\begin{figure}\centering
\includegraphics[height=4cm,width=7cm]{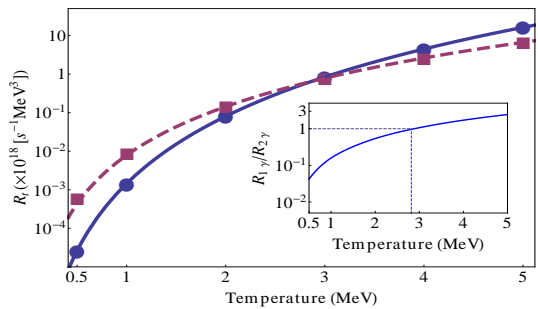}
\caption{\label{Fig:anniave} Temperature dependence of the total rates per volume for single-photon annihilation (blue dots) and two-photon annihilation (purple squares) in an $e^+e^-$ plasma. 
The inset displays the ratio of both annihilation channels. }
\end{figure}

As Fig.\,\ref{Fig:anniave} shows, the single-photon annihilation via the triple interaction \eqref{SPA}
can be sizable in a relativistic $e^+e^-$ plasma. The reason is that it exhibits a dependence on the particle density which is by one order higher than for the binary $e^+e^-$ annihilation into two photons. The steeper density scaling can compensate the additional factor of $\alpha$ involved in the process. The ratio of the two annihilation channels is of the order of $R_{1\gamma}/R_{2\gamma}\sim\alpha\rho/ m^3$. In the weakly relativistic case with $T\approx m$, this ratio reaches several percent. In this domain collective effects are still marginal since $\rho|\mathbf{k}_{\rm vir}|^{-3}\sim 0.1$. 

The rate ratio becomes unity at $T\approx 3$\,MeV, corresponding to a density $\rho\approx T^3\approx 6.5\times10^{32}\,$cm$^{-3}$ \cite{Thoma}. While the onset of collective effects in this regime might slightly modify rate predictions based on Eq.\,\eqref{annitotalrate},
our results clearly indicate the relevance of reaction \eqref{SPA} for a proper description of the annihilation dynamics in a relativistic $e^+e^-$ plasma. An analogous conclusion was drawn for the triple interactions considered in \cite{Ruffini}.

Note for comparison that three-photon annihilation, $e^+e^-\to 3\gamma$, will remain suppressed by a relative factor $\alpha$ with respect to two-photon annihilation since both processes share the same density dependence. Furthermore we notice that triple reactions analogous to \eqref{SPA} have also been found important in quark-gluon plasmas \cite{QCD}.

{\emph{Relation with atomic physics.}---}
Before proceeding to the conclusion, we note that reaction (\ref{SPA}) exhibits an interesting analogy in atomic physics. Free electrons typically recombine with ions via photo-emission (inverse photo-effect). At high electron densities, however, the radiationless channel of three-body recombination dominates where the recombining electron transfers its energy excess to a nearby partner electron \cite{Beiersdorfer}. This channel is proportional to the square of the electron density [cp. Eq.\,\eqref{Rpos}]. Single-photon annihilation in the presence of a second electron may be viewed as three-body recombination with the QED vacuum where one of the electrons recombines with a vacancy in the ``Dirac sea'' of negative-energy states. In both processes, the recoil absorbed by the assisting electron reduces the number of emitted photons by one, as compared to radiative recombination and two-photon annihilation, respectively.

{\emph{Conclusion.}---}
The triple reaction $e^+e^- + e \to \gamma + e'$ was studied in the relativistic domain, applying a novel theoretical approach. Under certain initial conditions the emitted photon exhibits very characteristic polarization properties and angular distributions, which offer an intuitive picture of this rather involved QED process and also might facilitate its detection in a dedicated experiment. Single-photon annihilation was shown to be important in relativistic $e^+e^-$ plasmas in equilibrium due to their extremely high density, which renders multi-particle correlation effects prominent.

{\emph{Acknowledgement.}---}
We thank K. Z. Hatsagortsyan and C. H. Keitel for useful discussions.
H.~H. acknowledges support from the National Basic Research Program of China 973 Program under Grant No. 2007CB815105.


\begin{thebibliography}{100}
\bibitem{Coulomb-effects}
Note that our results apply to particle energies slightly above the rest mass, $E^\pm \gtrsim 1.001m$, as we do not account for Coulomb corrections in the initial state.
\end{thebibliography}
\end{document}